\documentclass[9pt, conference]{IEEEtran}

\title{Large Language Model for Verilog Generation with \\
Code-Structure-Guided Reinforcement Learning}

\author{
Ning Wang$^{1}$ \quad Bingkun Yao$^{1}$ \quad Jie Zhou$^{2}$ \quad Yuchen Hu$^{2}$ \quad Xi Wang$^{2}$ \quad Nan Guan$^{1}$ \quad Zhe Jiang$^{2}$ \\[0.5em]
$^{1}$City University of Hong Kong \quad $^{2}$Southeast University
}

\usepackage{adjustbox}    
\usepackage{amsmath}
\usepackage{xspace}
\usepackage{enumitem}
\usepackage{algorithmic}
\usepackage{booktabs}
\usepackage{multirow}
\usepackage{xcolor}
\usepackage{threeparttable}
\usepackage{colortbl}
\usepackage{url}
\usepackage{amsfonts}
\usepackage{graphicx}
\usepackage{xspace}
\newcommand{\hide}[1]{}
\newcommand{\yhat}{\mathbf{\hat{y}}} 
\newcommand{\ours}{\emph{VeriSeek}}
\newcommand{\ourdataset}{\emph{VeriCores}}
\newcommand{\pass}{\emph{pass@}}

\newcommand{\hitfive}{\emph{hit@}5}

\usepackage{subcaption}
\usepackage[hidelinks]{hyperref}
\usepackage[linesnumbered,ruled,vlined]{algorithm2e}

\SetKwInput{KwInput}{Input}                
\SetKwInput{KwOutput}{Output}              
\SetNlSty{textbf}{\footnotesize}{:}

\AtBeginDocument{%
  }

\begin{document}

\maketitle

\begin{abstract}
Recent advancements in large language models (LLMs) have sparked significant interest in the automatic generation of Register Transfer Level (RTL) designs, particularly using Verilog.
Current research on this topic primarily focuses on pre-training and instruction tuning, but the effectiveness of these methods is constrained by the limited availability of training data, as public Verilog code is far less abundant than software code. 
In particular, these methods struggle to effectively capture Verilog's \emph{parallel} code structures, which fundamentally differ from the imperative, sequential control flow typical in most software programming languages.
This paper introduces \ours{}, an LLM enhanced by reinforcement learning using a limited amount of high-quality training data to achieve high Verilog code generation performance.
Our reinforcement learning approach employs code structure information as feedback signals to refine the pre-trained model, enabling it to effectively learn important patterns from Verilog code with parallel structures.
Experiments show that \ours{} outperforms state-of-the-art methods across multiple benchmarks. 
We release \ours{}'s complete implementation framework, including the dataset, source code, and model weights, at \url{https://anonymous.4open.science/r/veriseek-6467}.
\end{abstract}

\section{Introduction}

Large language models (LLMs) have demonstrated promising capabilities in various software programming tasks, prompting hardware design researchers to explore their applications in hardware design processes. One key application is using LLM for automatic generation of Hardware Description Language (HDL) code, such as Verilog, from specifications written in natural language.

The primary challenge in utilizing LLMs for Verilog code generation is the scarcity of training data, as the available open-source Verilog code is limited in both quantity and quality. 
Despite recent efforts in data collection and synthesis \cite{thakur2024verigen, zhang2024mg, liu2024craftrtl}, the volume of training data is still inadequate 
(much fewer than data available for training LLM to generate code in software programming languages). Moreover, using commercial models like GPT for training data synthesis or augmentation can hinder model performance, as it may introduce biases from the source models, leading to performance degradation through recursive training effects \cite{shumailov2024aicollapse}.

In this work, we aim to explore effective methods to train LLMs for Verilog code generation using limited data. Typically, training coding-oriented LLMs involves three stages \cite{chen2021evaluating}. The first stage, \emph{pre-training}, utilizes vast corpora of code and documentation to let the model understand the fundamental programming concepts and syntax. The second stage, \emph{instruction tuning}, enhances the model's ability to interpret and execute specific coding tasks. Finally, \emph{post-training}, typically using \emph{reinforcement learning}, adapts the model to specific programming paradigms. Comparing with pre-training and instruction tuning, post-training typically requires much less data as it emphasizes on exploring the model's existing capabilities rather than acquiring new information \cite{jiang2024survey}. Existing research on LLM training for Verilog code generation primarily focuses on the pre-training and instruction tuning stages \cite{thakur2024verigen, zhao2024codev, liu2024rtlcoder}. In contrast, our work emphasizes on the post-training stage. Specifically, we apply reinforcement learning to aggressively explore the parameter space and achieve better learning performance with limited training data.
\begin{figure}[h!]
\centering
\includegraphics[width=\linewidth]{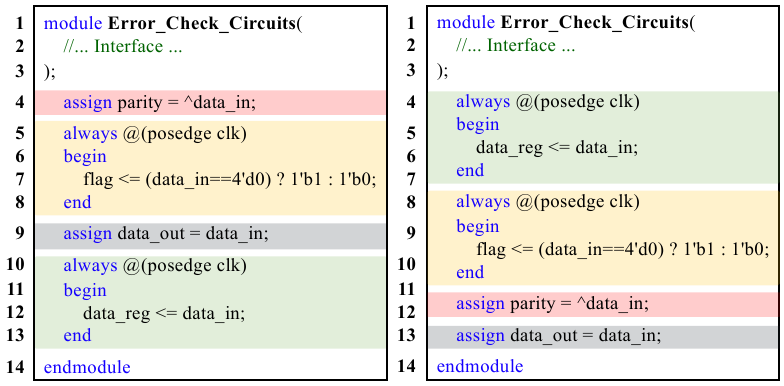}
    \caption{Two functionally equivalent Verilog modules with different token sequences. The left implementation follows (\footnotesize$parity \rightarrow flag \rightarrow data_{out} \rightarrow data_{reg}$\normalsize) sequence, whereas the right one is (\footnotesize$data_{reg} \rightarrow flag \rightarrow parity \rightarrow data_{out}$\normalsize). Corresponding colors between the left and right implementations represent identical code segments.
    }
\label{fig:example}
\end{figure}

However, applying reinforcement learning to post-train LLMs for Verilog code generation presents significant challenges. Although reinforcement learning has proven effective for post-training LLMs for \emph{software} code generation \cite{le2022coderl,liu2023rltf}, it performs poorly when directly applied to Verilog code generation (Section \ref{sec:exp} provides detailed experimental results illustrating this). A primary challenge stems from Verilog's inherent \emph{parallel} structures, which contrast with the \emph{sequential} execution typical of most software programming languages. For instance, Fig. \ref{fig:example} showcases two Verilog code segments that are functionally identical but exhibit substantial differences if compared as token sequences. 

This work introduces \ours{}, an LLM developed from DeepSeekCoder \cite{deepseekcoder} and enhanced by reinforcement learning with a novel reward function to address the aforementioned challenge. Our reward function assesses the generated code by comparing 
its structural similarities with the reference code, enabling the model to effectively capture Verilog-specific code patterns. Specifically, we convert the code into an Abstract Syntax Tree (AST) \cite{baxter1998clone} and develop a similarity scoring algorithm to evaluate the structural correspondence. This reward function is integrated with the Proximal Policy Optimization (PPO) algorithm \cite{ppo} to post-train the model. \ours{} outperforms existing state-of-the-art models \cite{zhang2024mg, goh2024english, thakur2024verigen, liu2024rtlcoder} on Verilog code generation benchmarks, RTLLM2.0 \cite{rtllm2} and VerilogEval \cite{verilogeval}.


\begin{figure*}[h!]
\centering
    \includegraphics[width=0.9\linewidth]{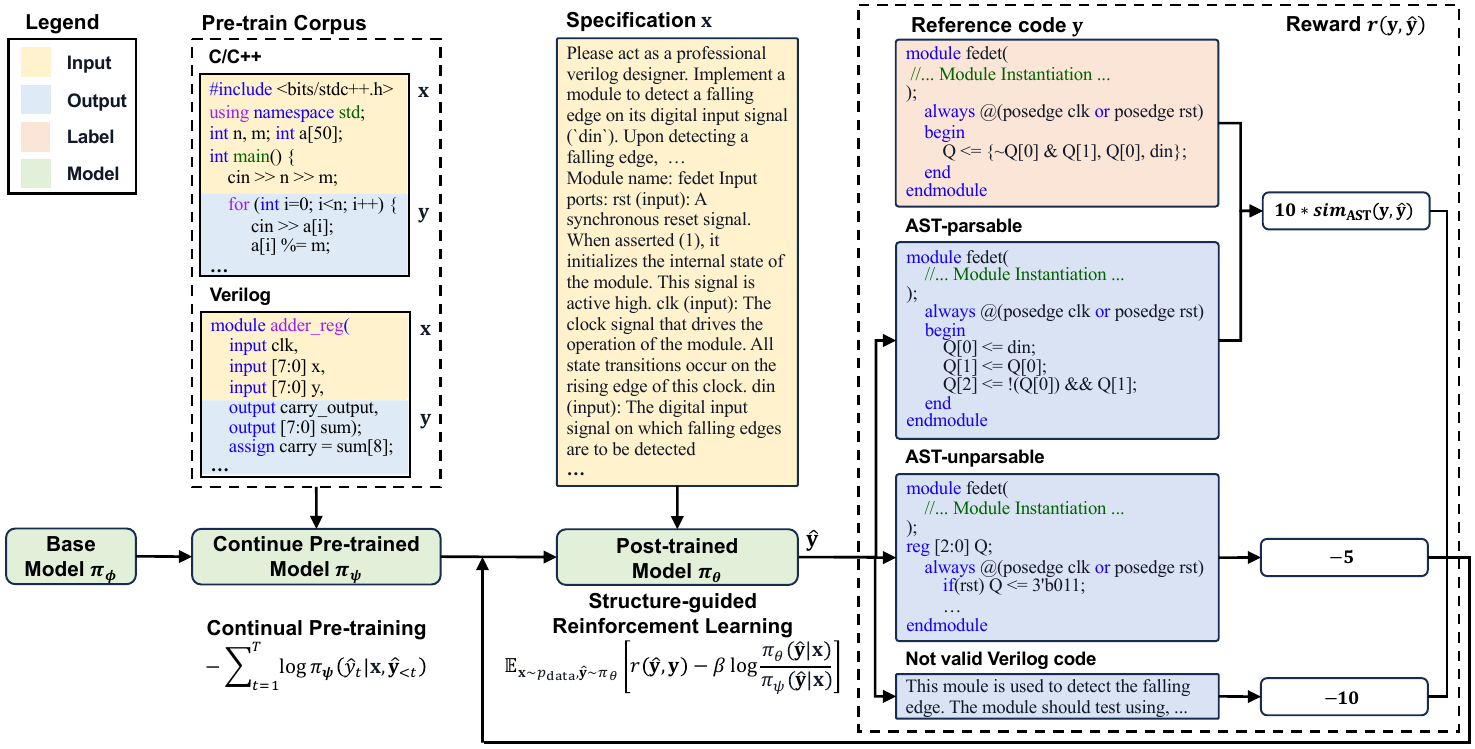}
    \caption{
    Overview of \ours{}'s training pipeline and reward mechanism.  
    Starting from a base model $\pi_\phi$, the model is trained on Verilog and C/C++ code to get $\pi_{\psi}$. 
    In the subsequent reinforcement learning stage, the model $\pi_\theta$ learns to generate Verilog code $\yhat$ from natural language specifications $\mathbf{x}$ by optimizing a code-structure-guided reward function $r(\mathbf{y}, \yhat)$. 
    This reward function evaluates the similarity between generated and reference code using AST-based similarity $sim_{\mathrm{AST}}$. 
    For unparsable generations, negative rewards (-10 or -5) are assigned based on the severity of syntax violations, encouraging the model to maintain proper Verilog syntax and semantics. 
    }
\label{fig:overview}
\end{figure*}

While post-training does not rely on a large amount of training data, it is sensitive to data quality. This is because post-training employs more explicit and targeted training objectives, so the noise in training data may cause greater disruption to the model. \cite{gururangan2020don}. 
Therefore, we have curated a dataset \ourdataset{} derived from OpenCores \cite{opencores}, a repository recognized for its high-quality open-source hardware designs. Each instance in 
\ourdataset{} comprises a natural-language specification as the model input and a high-quality reference Verilog code. Both \ours{} and \ourdataset{} are released at \url{https://anonymous.4open.science/r/veriseek-6467}.



\section{Related Work}

\subsection{LLM for Verilog Code Generation}
Many studies have advanced LLM-based Verilog code generation.
Thakur \cite{thakur} contributed to Verilog code generation collection through synthetic data generation and repository preprocessing. \cite{liu2024rtlcoder} developed RTLCoder that outperforms GPT-3.5 by training on automatically generated datasets using GPT. MG-Verilog \cite{zhang2024mg} constructed a multi-grained dataset that pairs Verilog code with descriptions at different detail levels to improve the model's instruction-following capability. BetterV \cite{pei2024betterv} introduced BetterV, which creates training datasets by converting Verilog code to C language, enabling LLMs to leverage their knowledge of general-purpose programming languages. Despite these efforts, the available datasets remain insufficient for comprehensive model training, which requires effective usage of data.


\subsection{Post-training LLMs for Coding}
Recent research has explored reinforcement learning approaches to improve LLMs' coding capabilities, specifically focusing on reward design mechanisms. \cite{chen2021evaluating} established the fundamental approach by using program outputs and runtime states to create execution-based reward signals. Subsequently, \cite{le2022coderl} developed a hierarchical reward framework that separates code evaluation into structural correctness and functional completion components, thus enabling more specific learning signals. \cite{liu2023rltf} extended this line of work by implementing a test-based feedback mechanism, where automatically generated test cases function as reward signals for comprehensive code evaluation. \cite{li2024ircoco} enhanced the reward signals by integrating static analysis metrics to address both functionality and code quality. Furthermore, \cite{dou2024stepcoder} implemented a compiler-feedback mechanism as reinforcement signals, allowing the model to learn from syntax errors and improve code generation iteratively. However, post-training LLMs for Verilog code generation with reinforcement learning remains unexplored.

\section{\ours{}}\label{sec:method}
\ours{} receives a natural-language specification as its input and outputs the corresponding Verilog code. As shown in Fig. \ref{fig:overview}, \ours{} is obtained on the base model through two training steps. The first step is continual pre-training, which enhances the LLM's basic understanding of Verilog syntax. The second step is reinforcement learning, which enables the LLM to learn Verilog-specific code patterns through iterative feedback and optimization.
For reinforcement learning, we design a code-structure-guided reward function that evaluates AST similarities between generated and reference code. Since this reward mechanism requires reliable reference code, we curate a high-quality dataset named \ourdataset{} and integrate these components into a PPO-based post-training framework.

\subsection{Continual Pre-training}
We use the public dataset VGen \cite{thakur2024verigen} for unsupervised continual pre-training. VGen aggregates Verilog repositories from GitHub and applies systematic filtering to remove duplicates. VGen also includes text extracted from $70$ Verilog textbooks. In total, VGen dataset contains approximately $50$ million tokens, with an $8:2$ ratio between Verilog code and natural language docstrings and comments.

Our experiments show that training with C/C++ code helps the model better understand and generate Verilog code, which aligns with results from previous research \cite{betterv}. Consequently, we expanded the training data withCodeSearchNet \cite{codesearchnet}, providing approximately $10$ million tokens with a $9$:$1$ ratio between C/C++ code and their documentations. The effectiveness of integrating C/C++ code in continual pre-training is evaluated in Section \ref{sec:exp}.

\subsection{Code-Structure-Guided Reinforcement Learning}
\subsubsection{Code-Structure-Guided Reward}

We first use Pyverilog \cite{pyverilog}, an open-source hardware design processing toolkit for Verilog, to generate AST of the code.
We then generate the \emph{cleaned AST},  by keeping the syntactic structure like operator types, module hierarchy and statement types
while discarding variable names and constant values from the original AST. The award function is calculated using 
$sim_{\mathrm{AST}}$, which compares the similarity of the cleaned ASTs of the generated code and
the reference code, as shown in Alg. \ref{algo:sim_ast}.

\begin{algorithm}[t]
\small

\KwInput{$t_1$ and $t_2$, the root nodes of two cleaned ASTs}
\KwOutput{Similarity in $[0.0, 1.0]$}


\If{$t_1$ and $t_2$ have the same type}
{
    $C_1 \leftarrow$ the set of $t_1$'s children nodes \\
    $C_2 \leftarrow$ the set of $t_2$'s children nodes \\
    $(sum, seen) \leftarrow (0, \emptyset)$ \\
    \For{every $c_1$ in $C_1$}
    {  
        $(best\_s, best\_c)\leftarrow (0, null)$ \\
        \For{every $c_2$ in $C_2 \setminus seen$}
        { 
            \If{$c_1$ and $c_2$ have the same type}
            {   
                $s \leftarrow sim_{\mathrm{AST}}(c_1, c_2)$ \\
                \If{$s > best\_s$}
                {
                    $(best\_s, best\_c) \leftarrow (s, c_2)$
                }
            }
        }
        \If{best\_c is not null}
        {
            $sum\leftarrow sum + best\_s$ \\
            $seen\leftarrow seen \bigcup \{best\_c\}$
            
        }
    }
    $max\_size \leftarrow \max\left(|c_1|, |c_2|\right)$ \\
    \If{$max\_size > 0$}
    {
        \KwRet{$sum/max\_size$}
    }
    \Else
    {
        \KwRet{1.0}
    }
}
\Else
{
    \KwRet{0.0}
}
\caption{$sim_{\mathrm{AST}}$: compute structural similarity between two cleaned ASTs}
\label{algo:sim_ast}
\end{algorithm}

$sim_{\mathrm{AST}}$ computes the similarity between two cleaned ASTs
through recursive comparison of their nodes and structures. 
The algorithm receives the root nodes $t_1, t_2$ of the two cleaned ASTs as the input. First, it checks whether $t_1$ and $t_2$
share the same type $(\text{Line \#}[1])$.
If yes, their children nodes are put into 
sets $C_1$ and $C_2$ respectively $(\text{Line \#}[2\text{-}3])$. Otherwise the similarity is $0.0$ $(\text{Line \#}[20\text{-}21])$ since these two cleaned ASTs with different types of root nodes are substaintially different. 


Then we iterates through each child node $c_1$ of $t_1$ to find its optimal match among $t_2$'s unmatched children $(\text{Line \#}[5])$. For each child $c_1$, it examines each unmatched child $c_2$ of $t_2$ $(\text{Line \#}[7])$. Here, $seen$ is the set of nodes in $C_2$ 
that have been matched.
If $c_1$ and $c_2$ have the same type, the algorithm computes their similarity through a recursive call to itself $(\text{Line \#}[9])$. If the result $s$ exceeds the current best similarity $best\_s$, the algorithm updates both $best\_s$ and $best\_c$ accordingly $(\text{Line \#}[10\text{-}11])$. 

The algorithm accumulates the similarity scores of matched child pairs $(\text{Line \#}[13])$ while tracking matched nodes in $seen$. These matched nodes are excluded from consideration for remaining $c_1$ comparisons $(\text{Line \#}[14])$ to ensure one-to-one matching.

The final similarity score is normalized to ensure the similarity is in the range of $[0.0, 1.0]$
We first set the maximum number of children nodes to $max\_size$ $(\text{Line \#}[15])$. If it is greater than 0 $(\text{Line \#}[16])$, indicating that at least one tree contains child nodes, the algorithm calculates the average of the summed similarities $(\text{Line \#}[17])$.
Alternatively, when both trees reach their leaf nodes with matching types $(\text{Line \#}[18])$, the algorithm returns the maximum similarity $1.0$ $(\text{Line \#}[19])$.

$sim_{\mathrm{AST}}$ is used to calculates the structural similarity between the cleaned AST of the generated code and the reference code in normal cases. There are also cases where the generated code failed to be parsed into an AST, for which we 
give a negative reward for punishment. In some cases, the LLM does not generate any valid Verilog code at all (e.g., the LLM just continues to write the specification instead of generating code), for which we give an even larger punishment. In our implementation, the reward is finally defined as:
\begin{equation*}
    \footnotesize
    r(\mathbf{y}, \yhat) =
    \begin{cases} 
    10 * {sim_{\mathrm AST}}(t_1, t_2), & \text{if } \mathbf{y} \text{ is AST-parsable} \\
    -5.0, & \text{if } \mathbf{y} \text{ is valid code but not AST-parsable} \\
    -10.0, & \text{if } \mathbf{y} \text{ is not valid code} 
    \end{cases}
    \label{eq:reward}
\end{equation*}
where $\mathbf{y}$ and $\yhat$ represents the generated code by the LLM and the reference code corresponding to the same specification; $t_1$ and $t_2$ represents the root node of $\mathbf{y}$ and $\yhat$, respectively.

\subsubsection{Proximal Policy Optimization}

We incorporate the reward introduced above into Proximal Policy Optimization (PPO) \cite{schulman2017proximal}, a widely-used reinforcement learning method,
to post-train our model. 

Here, we represent
the LLM as a policy (learned mapping function) $\pi_\theta$, where $\theta$ denotes the model parameters. This policy receives a design specification $\mathbf{x}$ and produces a text response $\yhat$ token by token:

\begin{equation}
\pi_\theta(\yhat\mid\mathbf{x}) = \prod_t \pi_\theta(\yhat_t\mid\mathbf{x}, \yhat_{<t}),
\end{equation}
 
PPO works by gradually improving the model's behavior through an iterative optimization process. During this process, the model learns from feedback while staying close to its original behavior.
Specifically, the objective function of PPO is defined as:

\begin{equation}
J_r(\pi_\theta) = \mathbb{E}_{\mathbf{x} \sim p_{\text{data}}, \yhat \sim \pi_\theta} \left[r(\yhat, \mathbf{y}) - \beta \log \frac{\pi_\theta(\yhat\mid\mathbf{x})}{\pi_{\psi}(\yhat\mid\mathbf{x})}\right].
\end{equation}

This objective comprises the code-structure-guided reward function $r(\yhat, \mathbf{y})$ that evaluates the quality of generated response, and a Kullback-Leibler (KL) divergence term weighted by $\beta$. The KL divergence measures how different the updated model is from the continual pre-trained model $\pi_{\psi}$ (the original continual pre-trained model).
The expectation is computed over inputs sampled from the data distribution and outputs from the current policy. Therefore, this conservative update strategy maintains the LLM's basic language capabilities while improving its Verilog generation performance.

\subsection{VeriCores Dataset}
\begin{figure}
\centering
    \includegraphics[width=\linewidth]{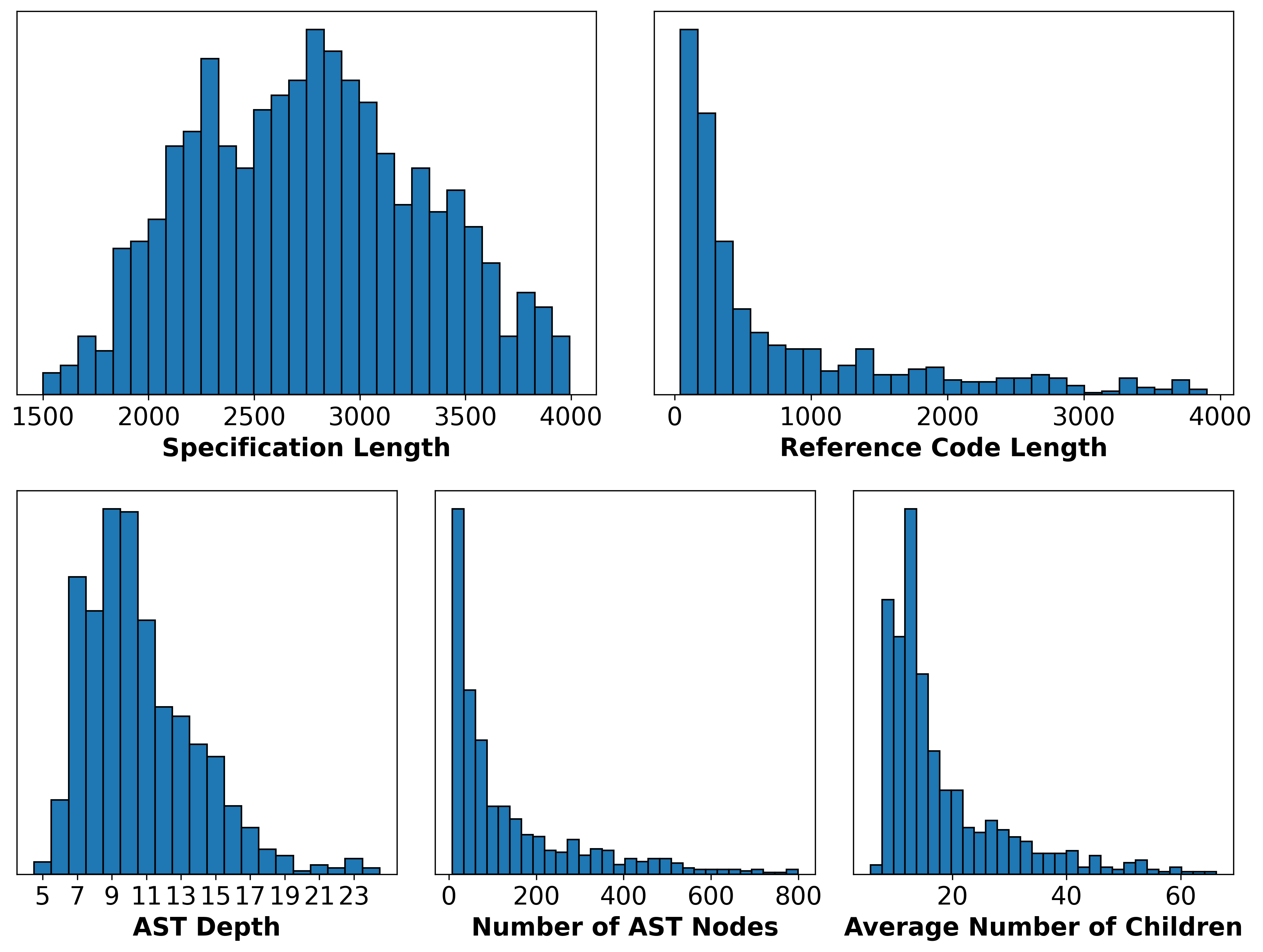}
    \caption{Statistics of the \ourdataset{} dataset, showing specification and code lengths, AST depth, node count, and branching factor metrics.}
\label{fig:dataset}
\end{figure}
Post-training's effectiveness depends mainly on data quality rather than quantity, as it employs targeted training objectives for domain adaptation. 
Our dataset \ourdataset{} collects high-quality specification-code pairs collected from Opencores \cite{opencores}, an open-source digital hardware development community. 
After filtering out instances where specifications or output code exceeded $4096$ tokens to align with LLMs' context window constraints and maintain consistent training quality, and removing instances with reference code failing AST parsing, the final dataset contains approximately $800$ instances.

Figure \ref{fig:dataset} presents statistics of \ourdataset{}, highlighting its diverse structural characteristics. The instruction lengths range from $1501$ to $3994$ tokens (mean: $2761.75$). The reference code ranges from $43$ to $3903$ tokens (mean: $784.61$). The AST structures vary in depth from $5$ to $24$ levels (mean: $10.68$), with node numbers between $8$ and $799$ (mean: $141.52$). The branching factor (the number of children of each node) ranges from $5.67$ to $66.21$ (mean: $18.01$).

\begin{table*}[h!]
\centering
\caption{Comparison of model performance on RTLLM2.0 and VerilogEval benchmarks, showing syntax and functional correctness metrics (\pass$1$, \pass$5$, \hitfive) for our models against open-source SOTA models, GPT-3.5 and GPT-4. All metrics are in \%.}
\label{tab:result}
\begin{threeparttable}
\setlength{\tabcolsep}{1pt}
\renewcommand\arraystretch{1.1}
\small
\begin{tabular}{cc cccccc cccccc}
\toprule
& & \multicolumn{6}{c}{\textbf{RTLLM2.0}} & \multicolumn{6}{c}{\textbf{VerilogEval}} \\
\cmidrule(lr){3-8} \cmidrule(lr){9-14}
& & \multicolumn{3}{c}{\textbf{Syntax}} & \multicolumn{3}{c}{\textbf{Function}} & \multicolumn{3}{c}{\textbf{Syntax}} & \multicolumn{3}{c}{\textbf{Function}} \\
\cmidrule(lr){3-8} \cmidrule(lr){9-14}
\multirow{-3}{*}{\textbf{Type}} &
\multirow{-3}{*}{\textbf{Model}} &
\pass${1}$ &
\pass${5}$ &
\hitfive &
\pass${1}$ &
\pass${5}$ &
\hitfive &
\pass${1}$ &
\pass${5}$ &
\hitfive &
\pass${1}$ &
\pass${5}$ &
\hitfive
\\ \midrule

& GPT-3.5 & 74.8 & 90.6 & 98.0 & 34.4 & 49.8 & 52.1 & 75.4 & 86.0 & 87.5 & 46.7 & 69.1 & 71.3 \\
\multirow{-2}{*}{\textbf{Closed-Source}} & GPT-4 & 80.0 & 89.5 & 98.9 & 47.9 & 58.0 & 68.9 & 76.1 & 86.8 & 87.4 & 60.0 & 70.6 & 72.8 \\ \midrule

& Goh-7B & 62.2 & 78.4 & 84.9 & 19.2 & 20.1 & 23.7 & 56.7 & 65.1 & 67.4 & 40.6 & 48.4 & 54.4 \\
& Thakur-16B & \colorbox{gray!25}{83.2} & 91.3 & 93.4 & 17.4 & 24.6 & 27.8 & 84.7 & 87.2 & 87.6 & 44.0 & 52.6 & 58.3 \\
& MG-Verilog-7B & 39.1 & 47.5 & 50.0 & 20.4 & 34.2 & 39.7 & 62.9 & 70.4 & 71.1 & 52.7 & 58.5 & 60.9 \\
\multirow{-4}{*}{\textbf{Open-Source}} & RTLCoder-7B & 73.4 & 89.7 & 91.3 & \colorbox{gray!25}{32.6} & 48.7 & 50.8 & \colorbox{gray!25}{86.6} & 97.7 & 98.9 & 61.2 & 76.5 & 80.4 \\ \midrule

\textbf{Base Models} & DeepSeekCoder-6.7B & 72.7 & 88.1 & 88.8 & 26.5 & 36.3 & 42.7 & 73.7 & 84.5 & 86.6 & 54.1 & 63.8 & 65.9 \\ \midrule

& \ours{}$_{PT}$-6.7B & 65.6 & 89.3 & 84.1 & 26.2 & 48.9 & 49.2 & 72.9 & 84.1 & 84.9 & 53.3 & 63.5 & 65.2 \\
& \ours$_{PTwC}$-6.7B & 72.5 & 94.2 & 95.4 & 30.1 & 50.7 & 51.4 & 76.3 & 87.4 & 88.2 & 58.4 & 68.5 & 71.9 \\
\multirow{-3}{*}{\textbf{Ours}} & \ours$_{PTwC+RL}$-6.7B & 73.5 & \colorbox{gray!25}{94.8} & \colorbox{gray!25}{96.0} & 31.9 & \colorbox{gray!25}{54.2} & \colorbox{gray!25}{52.0} & 85.1 & \colorbox{gray!25}{98.3} & \colorbox{gray!25}{99.1} & \colorbox{gray!25}{61.6} & \colorbox{gray!25}{76.9} & \colorbox{gray!25}{81.7} \\ \bottomrule
\end{tabular}
\begin{tablenotes}\footnotesize
    \item[+] \colorbox{gray!25}{Gray} background represents the best metric (excluding GPT-4).
\end{tablenotes}
\end{threeparttable}
\end{table*}

\section{Experiments and Performance Evaluation}\label{sec:exp}
\subsection{Training Details}

Based on the base model DeepSeekCoder-6.7B \cite{deepseekcoder}, we develop three variants of our model with different training strategies. All three versions has the same model size of $6.7$B parameters. 

\begin{itemize}[noitemsep,topsep=0pt]
    \item $\ours_{PT}$: Pre-trained with Verilog code only.
    \item $\ours_{PTwC}$: Pre-trained with both Verilog and C/C++ code.
    \item $\ours_{PTwC+RL}$: $\ours_{PTwC}$ post-trained by reinforcement learning.
\end{itemize}


Experiments are conducted on a server equipped with 8 A800-80G GPUs. All experiments utilize a cosine learning rate scheduler with a warmup phase comprising 10\% of the total training steps, and an AdamW optimizer \cite{adamw} with a weight decay of $0.05$. Additionally, we employ deepspeed ZeRO-3 offload \cite{deepspeed} for acceleration.

Following the hyper-parameter settings in the traning of the base model DeepSeekCoder, we adopt a peak learning rate of $1e^{-4}$ and a batch size of $32$ for continual pre-training, training for $1$ epoch. 
For reinforcement learning, we employ Low-rank Adaptation (LoRA) \cite{hu2021lora} on query and value projection matrices to reduce memory usage and training time for PPO's iterative optimization process.
We set a peak learning rate of $1e^{-5}$, a batch size of $8$, and train for $10$ epochs, with maximum sequence length of $2048$ tokens and generation parameters of temperature $0.2$ and top-p $0.95$. The duration of continual pre-training is approximately $1$ hour, whereas the reinforcement learning task requires about $1$ day to complete training. 
Reinforcement learning takes considerably longer time to converge than continual pre-training, primarily because of PPO's iterative update mechanism. 
PPO conducts multiple forward passes to collect trajectories and performs multiple optimization steps.

\subsection{Metric and Benchmark}

\subsubsection{Metric} 
We evaluate the models using the widely-adopted \pass$k$ metric for code generation, which is the percentage of problems solved by using $k$ generated programs per problem  \cite{thakur}: 

\begin{equation}
    pass@k := \mathbb{E}_{i} \left[ 1 - \frac{\binom{n-c_i}{k}}{\binom{n}{k}} \right]
\end{equation}
where $n$ is the total number of trials for each specification and
$c_i$ is the number of correct code generations for task $i$. We set
$n=20$ in this experiment for comparison with baselines.
When any code within the $k$ trials successfully passes the test, we consider the task addressed. The \pass${k}$  metric therefore represents the estimated percentage of design tasks that can be successfully completed.
We measure syntax and functional \pass${1}$ and \pass${5}$ metrics, where `syntax' means that the code is compiled successfully and `functional' means that the code passes the testbench.

In \cite{liu2024rtlcoder}, RTLCoder was evaluated with a metric called `pass@5', 
which evaluates whether any test among $5$ trials passes the testbench. This metric differs from the above defined~\pass$k, k=5$ metric. To enable direct comparison with RTLCoder while avoiding confusion, we rename this metric as \hitfive{} and include it in our evaluation.

\subsubsection{Benchmark}
We conduct performance evaluation with two Verilog code generation benchmarks: RTLLM2.0 and VerilogEval.
RTLLM2.0 \cite{rtllm2} contains 50 design tasks in four categories: Arithmetic, Control, Memory and Miscellaneou. 
VerilogEval \cite{verilogeval} is a comprehensive benchmark with tasks ranging from simple combinational logic to complex state machines. Since we focus on natural language specifications, we exclude hand-written tasks in VerilogEval with specifications not in nature languages (e.g., using waveforms to describe the expected output).
For both RTLLM2.0 and VerilogEval, each design task has a specification and corresponding testbench. 
Following the testing methods in  \cite{liu2024rtlcoder}, we evaluate syntax and functional pass rate using ModelSim \cite{modelsim}. Syntax pass requires the generated code to be compiled successfully, while functional pass requires 
the code to succeed simulations with the testbench.

\subsection{Performance Evaluation}
As shown in Table \ref{tab:result}, our model demonstrates strong capabilities across both RTLLM2.0 and VerilogEval benchmarks. \ours{}$_{PT}$ and \ours{}$_{PTwC}$ show substantial improvements over the base model DeepSeekCoder-6.7B. After reinforcement learning, our final model \ours{}$_{PTwC+RL}$ achieves impressive results on both benchmarks.

\begin{figure}
\centering
    \includegraphics[width=0.85\linewidth]{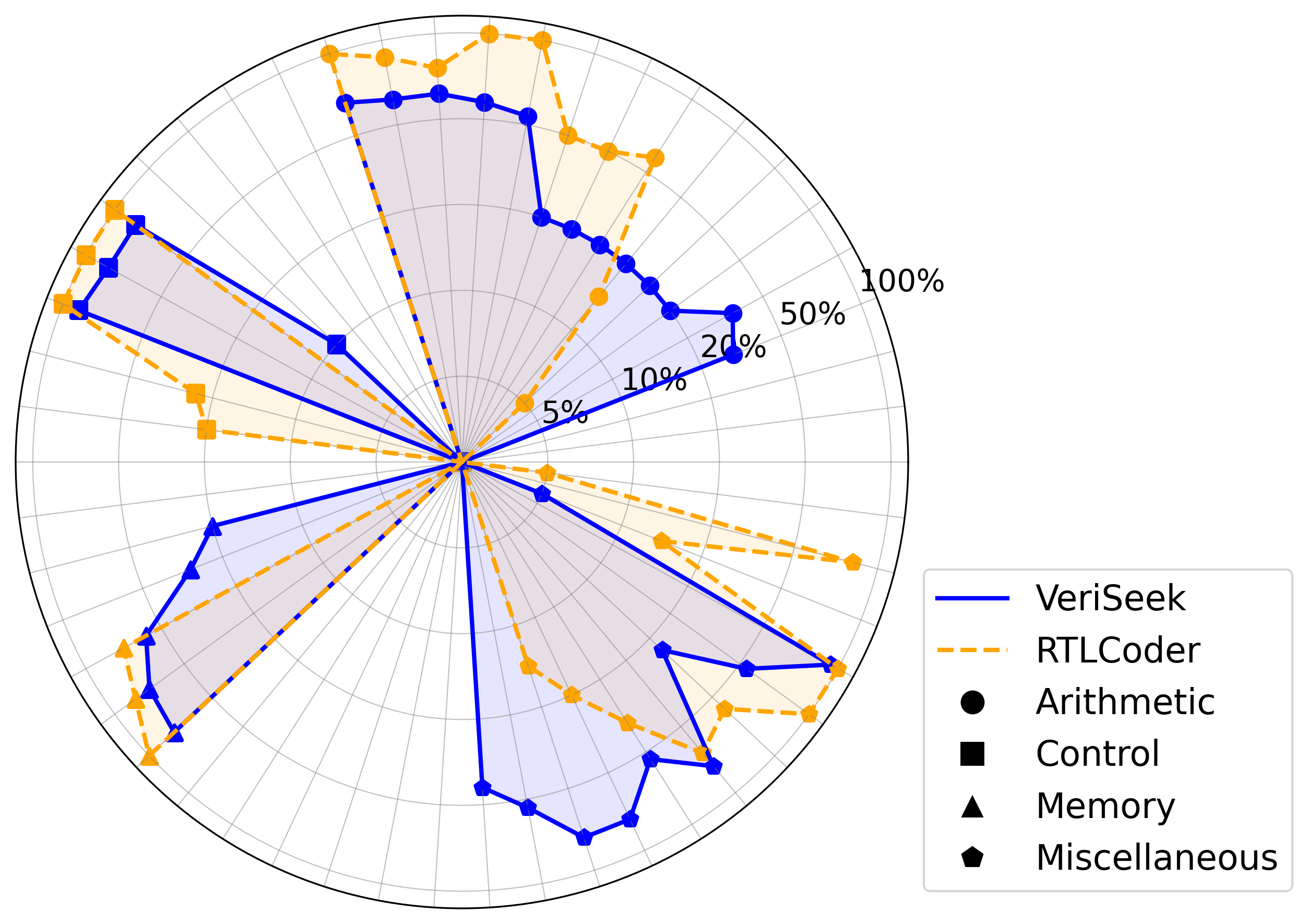}
    \caption{\pass$5$ performance comparison between VeriSeek and RTLCoder across different task categories in RTLLM2.0 benchmark.}
\label{fig:comparison}
\end{figure}

On RTLLM2.0, \ours{}$_{PTwC+RL}$ achieves the best performance among open-source SOTA models Thakur \cite{thakur}, ChipGPT \cite{chang2023chipgpt} and RTLCoder \cite{liu2024rtlcoder}. 
While GPT-4 maintains the overall best performance, our model surpasses GPT-3.5 across almost all metrics in both syntax and functional evaluations.
The radar chart in Fig.\ref{fig:comparison} illustrates the \pass{$5$} performance across different categories in RTLLM2.0, where each axis represents a specific task. While RTLCoder achieves higher performance on certain individual design tasks, \ours{}$_{PTwC+RL}$ demonstrates more consistent performance by successfully handling a broader range of tasks, particularly in Miscellaneous-related tasks.
On VerilogEval, \ours{}$_{PTwC+RL}$ achieves the best performance among all models, including GPT-4.
In particular, \ours{}$_{PTwC+RL}$ achieves functional pass rates of 61.6\% (\pass{$1$}), 76.9\% (\pass{$5$}) and 81.7\% (\hitfive), which notably exceeds GPT-4's performance of 60.0\% (\pass{$1$}), 70.6\% (\pass{$5$}) and 72.8\% (\hitfive).

While \ours{}$_{PTwC+RL}$ outperforms all existing open-source models and GPT-3.5, it may lag behind GPT-4 in certain design tasks, potentially due to GPT-4's substantially larger size (it's widely believed that GPT-4 largely exceed GPT-3's 175 billion parameters). The competitive performance of GPT-4 does not diminish the value of \ours{}$_{PTwC+RL}$. The hardware design domain is particularly sensitive to intellectual property protection and the security of designs. Consequently, hardware design companies may prefer deploying their own LLMs rather than relying on closed-source models like GPT-4.
While the long-term competition between open-source and closed-source models for Verilog code generation is likely to continue, our work advances the state-of-the-art of open-source models.

\subsection{Ablation Study}

\subsubsection{Instruction Tuning}
First we discuss our attempt of applying instruction tuning 
to the training of \ours{}.
Instruction tuning is a process where LLMs are trained on datasets comprising instructions and corresponding responses \cite{instructfinetune}, enhancing their ability to accurately follow human instructions. Instruction tuning employ Maximum Likelihood Estimation (MLE) to find the best parameters:
\begin{equation}
\mathcal{L}_{\text{mle}} = - \sum_{t=1}^T \log \pi_\psi ( \yhat_t \mid \mathbf{x}, \yhat_{<t})
\end{equation}
which measures how well the model predicts each token $\yhat_t$ given the instruction $\mathbf{x}$ and previous tokens $\yhat_{<t}$.

We instruction-tuned our continual pre-trained model on the Opencores dataset, denoted as \ours$_{PTwC+FT}$. Then we post-train \ours$_{PTwC+FT+RL}$ using PPO as the same settings with \ours$_{PTwC+RL}$.
As shown in Table \ref{tab:ablation}, the model achieves a slight improvement in the functional \pass${1}$ metric while performing poorly across other evaluation metrics. The degraded performance can be attributed to two factors. First, the exposure bias in auto-regressive sequence generation causes model deviations from reference code, as the model depends on its generated tokens rather than reference tokens for predictions \cite{liu2024rtlcoder, bengio2015scheduled}. Second, the sequential processing of auto-regressive generation conflicts with Verilog's inherent parallel structures, thereby limiting the model's ability to maintain consistent relationships between concurrent blocks and signals and resulting in poor generation diversity.

\begin{table}[h]
\centering
\caption{Ablation study on instruction tuning, learned reward by paired generations and parallel-unaware reward on RTLLM2.0.}
\label{tab:ablation}
\begin{threeparttable}
\setlength{\tabcolsep}{1.5pt}
\renewcommand\arraystretch{1.2}
\begin{tabular}{ccccccc}
\toprule
& \multicolumn{3}{c}{\textbf{Syntax}} & \multicolumn{3}{c}{\textbf{Function}} \\ \cline{2-7} 
\multirow{-2}{*}{\textbf{Model}} &
\pass${1}$ &
\pass${5}$ &
\hitfive &
\pass${1}$ &
\pass${5}$ &
\hitfive
\\ \specialrule{.08em}{.2em}{.2em} 

\ours$_{PTwC}$ & 72.5 & 94.2 & 95.4 & 30.1 & 50.7 & 51.4 \\
\ours$_{PTwC+FT}$ & 72.3 & 94.6 & 93.1 & \colorbox{gray!25}{33.1} & 46.1 & 48.3 \\
\ours$_{PTwC+FT+RL}$ & 69.7 & 89.6 & 91.2 & 32.4 & 45.7 & 47.5 \\
\midrule
\ours{}$_{PTwC+BT}$ & 38.4 & 50.2 & 54.6 & 19.5 & 33.1 & 38.9 \\
\ours$_{PTwC+SEQ}$ & 68.6 & 91.0 & 92.2 & 27.5 & 48.3 & 49.7 \\
\midrule
\ours$_{PTwC+RL}$ & \colorbox{gray!25}{73.5} & \colorbox{gray!25}{94.8} & \colorbox{gray!25}{96.0} & 31.9 & \colorbox{gray!25}{54.2} & \colorbox{gray!25}{52.0} \\ \bottomrule
\end{tabular}
\begin{tablenotes} 
\item[+] \colorbox{gray!25}{Gray} background represents the best metric.
\end{tablenotes}
\vspace{-5pt}
\end{threeparttable}
\end{table}

\subsubsection{Learn Reward from Paired Generations}

Now we discuss the attempt to apply post-training methods commonly used in natural language tasks to the training of $\ours{}$. In natural language tasks such as question answering, post-training of LLMs typically employs a learnable reward model $r$ with Bradley-Terry modeling 
$\frac{e^{r(\mathbf{x}, \mathbf{y}_w)}}{e^{r(\mathbf{x}, \mathbf{y}_w)} + e^{r(\mathbf{x}, \mathbf{y}_l)}}$
to capture human preferences between response pairs \cite{hfrl1, hfrl2}. The reward model is trained by minimizing:
\begin{equation}
\mathcal{L} = \mathbb{E}_{(\textbf{x},\textbf{y}_w,\textbf{y}_l) \sim \mathcal{D}} \left[ -\log \left(\frac{e^{r(\textbf{x}, \textbf{y}_w)}}{e^{r(\textbf{x}, \textbf{y}_w)} + e^{r(\textbf{x}, \textbf{y}_l)}}\right) \right]
\end{equation}

In this objective, the model learns to assign higher scores $r$ to winning responses compared to losing ones, where the exponential terms are normalized through softmax to obtain probabilities. 
To train the reward model, we construct dataset $\mathcal{D}$ containing triplets $(\mathbf{x}, \mathbf{y}_w, \mathbf{y}_l)$, where $\mathbf{x}$ represents the specification, $\mathbf{y}_w$ denotes the generated code that passes the benchmark, and $\mathbf{y}_l$ represents the failed generation.
After training the reward model, we apply PPO with the same settings as before and get the model $\ours{}_{PTwc+BT}$.

The experimental results in Table \ref{tab:ablation} show degraded performance compared even to the continual pre-trained baseline, as the Bradley-Terry model optimizes only relative differences between responses while ignoring absolute reward values. While this relative preference approach is suitable for aligning with general human values, it becomes problematic in coding tasks where the space of correct solutions is substantially smaller than that of incorrect ones.

\subsubsection{Effectiveness of the Code-Structure-Guided Reward}
To evaluate the effectiveness of code-structure-guided reward which is designed to the capture the parallel structure of Verilog code, we modify Alg. \ref{algo:sim_ast} by implementing sequential node comparison between two ASTs:
\begin{algorithm}[!ht]
\small
\caption{$sim_{\mathrm{AST\_SEQ}}$: compute sequential structures similarity between two cleaned ASTs}
\label{algo:sim_ast_seq}

\KwInput{$t_1$ and $t_2$, the root nodes of two Cleaned ASTs}
\KwOutput{Similarity in $[0.0, 1.0]$}
... \\
\If{$t_1$ and $t_2$ have the same type}
{   
    $sum \leftarrow 0$ \\
    \For{paired $(c_1, c_2)$ in $(C_1, C_2)$}
    {
        $sum \leftarrow sum + sim_{\mathrm{AST\_SEQ}}(c_1, c_2)$ \\
    }
    $max\_size \leftarrow \max(|c_1|, |c_2|)$ \\
    ...
}

... 

\end{algorithm}
This modification implements one-by-one comparison between corresponding children nodes between two ASTs. Then we use same settings as \ours$_{PTwc+RT}$ to post-train the continued pre-trained model. 
We refer to this variant as \ours$_{PTwc+SEQ}$. 
As shown in Table \ref{tab:ablation}, this does not  improve the performance, indicating the reward sequentially compares ASTs does is ineffective.

\subsubsection{Performance with Different Temperatures}

\begin{figure}[h!]
\centering
    \includegraphics[width=\linewidth]{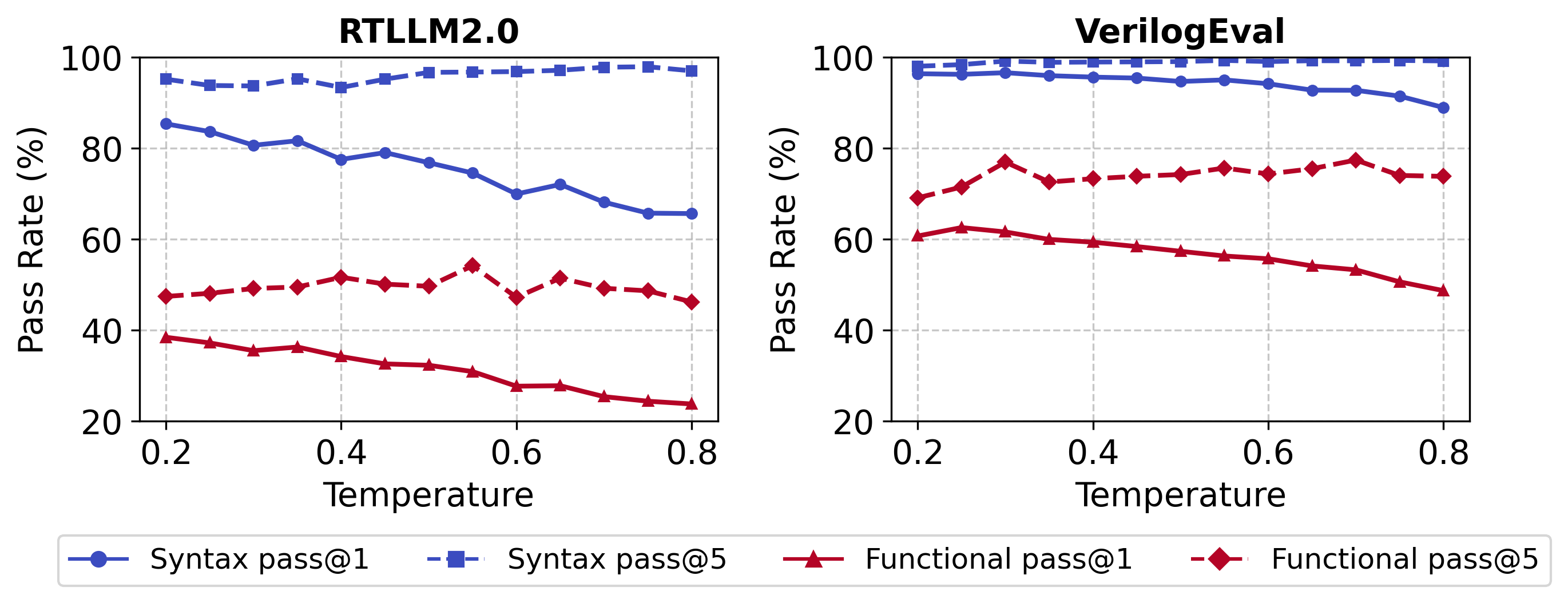}
    \caption{
    Temperature analysis of \ours{}$_{PTwC+RL}$.
    }
\label{fig:temperature}
\end{figure}

Fig. \ref{fig:temperature} presents the effect of sampling temperature on \ours{}$_{PTwC+RL}$ performance is evaluated across two benchmarks RTLLM2.0 and VerilogEval, with temperature ranging from $0.2$ to $0.8$ at intervals of $0.05$. The experimental results show that increasing temperature consistently degrades both syntax and functional \pass{$1$}. However, the syntax \pass{$5$}, functional \pass{$5$} metrics are stable across different temperatures.

\subsection{Training Dynamics of Reinforcement Learning}
\begin{figure}
\centering
    \includegraphics[width=0.9\linewidth]{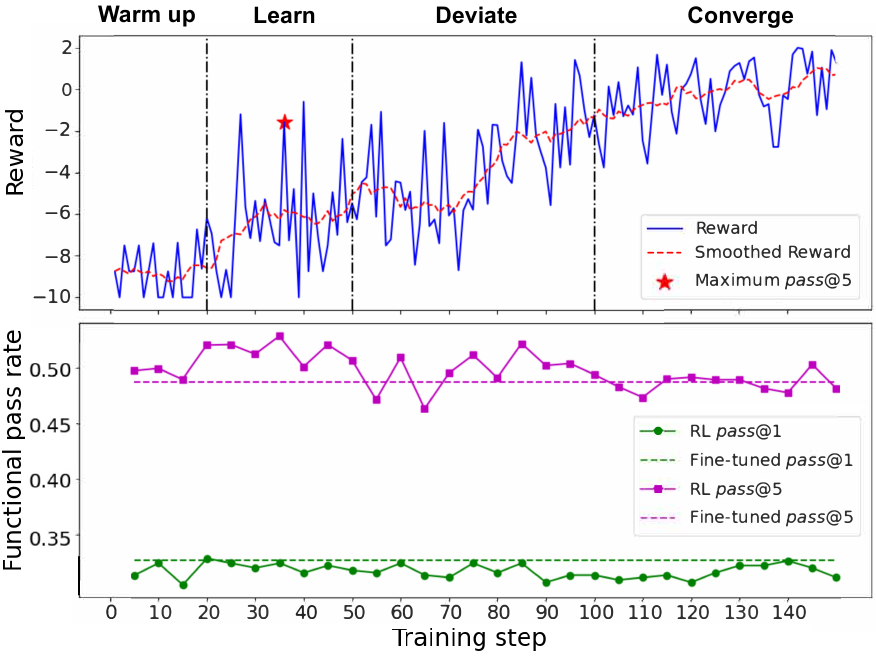}
    \caption{
    Reward, \pass{$1$} and \pass{$5$} on RTLLM2.0 during training.
    }
    \label{fig:ppo_reward}
\end{figure}

Fig.\ref{fig:ppo_reward} shows the reward and functional \pass${1}$ and \pass${5}$ every five training steps during reinforcement learning. The optimal model performance appears during the early training stages rather than at convergence. As training progresses, the model converges to a fine-tuned state. We can split the training process into four stages. 

\begin{figure}
\centering
    \includegraphics[width=0.8\linewidth]{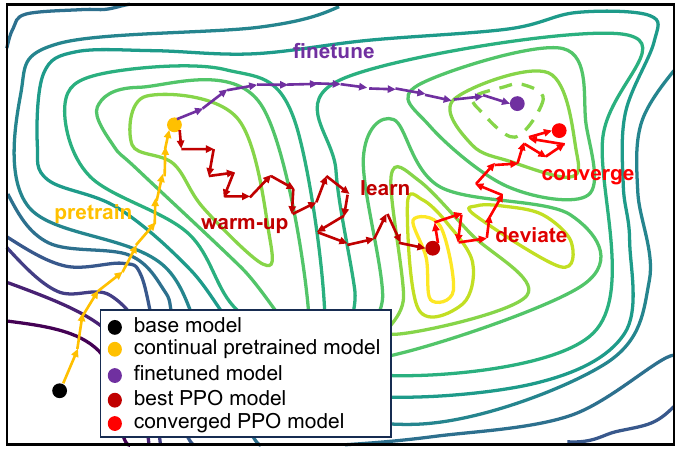}
    \caption{
    Optimization trajectory across different training stages. 
    }
\label{fig:dynamic}
\end{figure}

Fig.\ref{fig:dynamic} illustrates the learning dynamics across various stages. 
In the \emph{warm-up} stage (0-20 steps), starting from the bottom-left corner with a low pass rate, the model escapes from suboptimal solutions of the continual pre-trained model. In the \emph{learning} stage (20-50 steps), the code-structure-guided reward helps the model achieve better performance. The model walks towards the optimal area, the yellow region in Fig.\ref{fig:dynamic}.
The \emph{deviation} stage (50-100 steps), shows the model deviating from the optimal region due to misalignment between reward signals and evaluation metrics. 
During the \emph{convergence} stage (100-150 steps), the model achieves an instruction-tuned state, as illustrated by the light red trajectory in Fig. \ref{fig:dynamic}. This convergence occurs because AST comparison, while effective for parallel structures, cannot fully capture implementation requirements from specifications due to its dependence on reference code.

These training dynamics demonstrate that well-defined rewards that can capture the requirement of specification are essential for stabilizing the reinforcement learning.

\section{Conclusion}
We presented \ours{}, a reinforcement learning approach for post-training LLMs in Verilog code generation using structure-guided rewards. By leveraging AST-based structural similarity analysis, \ours{} effectively addresses the challenge of limited training data in Verilog generation. Our experimental results show that \ours{} achieves state-of-the-art performance on standard benchmarks, surpassing GPT-4 on VerilogEval. The approach specifically focus on Verilog's parallel structures, which differ from sequential software codes. Future work could focus on developing reward functions that better capture Verilog's parallel execution patterns and reinforcement learning strategies tailored for hardware design characteristics.

\clearpage

\bibliographystyle{abbrv}
\bibliography{ref}

\begin{thebibliography}{10}

\bibitem{baxter1998clone}
I.~D. Baxter, A.~Yahin, L.~Moura, M.~Sant'Anna, and L.~Bier.
\newblock Clone detection using abstract syntax trees.
\newblock In {\em Proceedings. International Conference on Software Maintenance (Cat. No. 98CB36272)}, pages 368--377. IEEE, 1998.

\bibitem{bengio2015scheduled}
S.~Bengio, O.~Vinyals, N.~Jaitly, and N.~Shazeer.
\newblock Scheduled sampling for sequence prediction with recurrent neural networks.
\newblock {\em Advances in neural information processing systems}, 28, 2015.

\bibitem{chang2023chipgpt}
K.~Chang, Y.~Wang, H.~Ren, M.~Wang, S.~Liang, Y.~Han, H.~Li, and X.~Li.
\newblock Chipgpt: How far are we from natural language hardware design.
\newblock {\em arXiv preprint arXiv:2305.14019}, 2023.

\bibitem{chen2021evaluating}
M.~Chen, J.~Tworek, H.~Jun, Q.~Yuan, H.~P. D.~O. Pinto, J.~Kaplan, H.~Edwards, Y.~Burda, N.~Joseph, G.~Brockman, et~al.
\newblock Evaluating large language models trained on code.
\newblock {\em arXiv preprint arXiv:2107.03374}, 2021.

\bibitem{dou2024stepcoder}
S.~Dou, Y.~Liu, H.~Jia, L.~Xiong, E.~Zhou, W.~Shen, J.~Shan, C.~Huang, X.~Wang, X.~Fan, et~al.
\newblock Stepcoder: Improve code generation with reinforcement learning from compiler feedback.
\newblock {\em arXiv preprint arXiv:2402.01391}, 2024.

\bibitem{goh2024english}
E.~Goh, M.~Xiang, I.~Wey, T.~H. Teo, et~al.
\newblock From english to asic: Hardware implementation with large language model.
\newblock {\em arXiv preprint arXiv:2403.07039}, 2024.

\bibitem{deepseekcoder}
D.~Guo, Q.~Zhu, D.~Yang, Z.~Xie, K.~Dong, W.~Zhang, G.~Chen, X.~Bi, Y.~Wu, Y.~Li, et~al.
\newblock Deepseek-coder: When the large language model meets programming--the rise of code intelligence.
\newblock {\em arXiv preprint arXiv:2401.14196}, 2024.

\bibitem{gururangan2020don}
S.~Gururangan, A.~Marasovi{\'c}, S.~Swayamdipta, K.~Lo, I.~Beltagy, D.~Downey, and N.~A. Smith.
\newblock Don't stop pretraining: Adapt language models to domains and tasks.
\newblock {\em arXiv preprint arXiv:2004.10964}, 2020.

\bibitem{hu2021lora}
E.~J. Hu, Y.~Shen, P.~Wallis, Z.~Allen-Zhu, Y.~Li, S.~Wang, L.~Wang, and W.~Chen.
\newblock Lora: Low-rank adaptation of large language models.
\newblock {\em arXiv preprint arXiv:2106.09685}, 2021.

\bibitem{codesearchnet}
H.~Husain, H.-H. Wu, T.~Gazit, M.~Allamanis, and M.~Brockschmidt.
\newblock Codesearchnet challenge: Evaluating the state of semantic code search.
\newblock {\em arXiv preprint arXiv:1909.09436}, 2019.

\bibitem{jiang2024survey}
J.~Jiang, F.~Wang, J.~Shen, S.~Kim, and S.~Kim.
\newblock A survey on large language models for code generation.
\newblock {\em arXiv preprint arXiv:2406.00515}, 2024.

\bibitem{le2022coderl}
H.~Le, Y.~Wang, A.~D. Gotmare, S.~Savarese, and S.~C.~H. Hoi.
\newblock Coderl: Mastering code generation through pretrained models and deep reinforcement learning.
\newblock {\em Advances in Neural Information Processing Systems}, 35:21314--21328, 2022.

\bibitem{li2024ircoco}
B.~Li, Z.~Sun, T.~Huang, H.~Zhang, Y.~Wan, G.~Li, Z.~Jin, and C.~Lyu.
\newblock Ircoco: Immediate rewards-guided deep reinforcement learning for code completion.
\newblock {\em Proceedings of the ACM on Software Engineering}, 1(FSE):182--203, 2024.

\bibitem{liu2023rltf}
J.~Liu, Y.~Zhu, K.~Xiao, Q.~Fu, X.~Han, W.~Yang, and D.~Ye.
\newblock Rltf: Reinforcement learning from unit test feedback.
\newblock {\em arXiv preprint arXiv:2307.04349}, 2023.

\bibitem{verilogeval}
M.~Liu, N.~Pinckney, B.~Khailany, and H.~Ren.
\newblock Verilogeval: Evaluating large language models for verilog code generation.
\newblock In {\em 2023 IEEE/ACM International Conference on Computer Aided Design (ICCAD)}, pages 1--8. IEEE, 2023.

\bibitem{liu2024craftrtl}
M.~Liu, Y.-D. Tsai, W.~Zhou, and H.~Ren.
\newblock Craftrtl: High-quality synthetic data generation for verilog code models with correct-by-construction non-textual representations and targeted code repair.
\newblock {\em arXiv preprint arXiv:2409.12993}, 2024.

\bibitem{liu2024rtlcoder}
S.~Liu, W.~Fang, Y.~Lu, Q.~Zhang, H.~Zhang, and Z.~Xie.
\newblock Rtlcoder: Outperforming gpt-3.5 in design rtl generation with our open-source dataset and lightweight solution.
\newblock In {\em 2024 IEEE LLM Aided Design Workshop (LAD)}, pages 1--5. IEEE, 2024.

\bibitem{rtllm2}
S.~Liu, Y.~Lu, W.~Fang, M.~Li, and Z.~Xie.
\newblock Openllm-rtl: Open dataset and benchmark for llm-aided design rtl generation.
\newblock 2024.

\bibitem{adamw}
I.~Loshchilov and F.~Hutter.
\newblock Decoupled weight decay regularization.
\newblock {\em arXiv preprint arXiv:1711.05101}, 2017.

\bibitem{opencores}
{OpenCores}.
\newblock Opencores.
\newblock \url{https://opencores.org/}, 2024.
\newblock Accessed: 2024-11-14.

\bibitem{hfrl2}
L.~Ouyang, J.~Wu, X.~Jiang, D.~Almeida, C.~Wainwright, P.~Mishkin, C.~Zhang, S.~Agarwal, K.~Slama, A.~Ray, et~al.
\newblock Training language models to follow instructions with human feedback.
\newblock {\em Advances in neural information processing systems}, 35:27730--27744, 2022.

\bibitem{pei2024betterv}
Z.~Pei, H.-L. Zhen, M.~Yuan, Y.~Huang, and B.~Yu.
\newblock Betterv: Controlled verilog generation with discriminative guidance.
\newblock {\em arXiv preprint arXiv:2402.03375}, 2024.

\bibitem{betterv}
Z.~Pei, H.-L. Zhen, M.~Yuan, Y.~Huang, and B.~Yu.
\newblock Betterv: Controlled verilog generation with discriminative guidance.
\newblock {\em arXiv preprint arXiv:2402.03375}, 2024.

\bibitem{deepspeed}
J.~Rasley, S.~Rajbhandari, O.~Ruwase, and Y.~He.
\newblock Deepspeed: System optimizations enable training deep learning models with over 100 billion parameters.
\newblock In {\em Proceedings of the 26th ACM SIGKDD International Conference on Knowledge Discovery \& Data Mining}, pages 3505--3506, 2020.

\bibitem{ppo}
J.~Schulman, F.~Wolski, P.~Dhariwal, A.~Radford, and O.~Klimov.
\newblock Proximal policy optimization algorithms.
\newblock {\em arXiv preprint arXiv:1707.06347}, 2017.

\bibitem{schulman2017proximal}
J.~Schulman, F.~Wolski, P.~Dhariwal, A.~Radford, and O.~Klimov.
\newblock Proximal policy optimization algorithms.
\newblock {\em arXiv preprint arXiv:1707.06347}, 2017.

\bibitem{shumailov2024aicollapse}
I.~Shumailov, Z.~Shumaylov, Y.~Zhao, N.~Papernot, R.~Anderson, and Y.~Gal.
\newblock Ai models collapse when trained on recursively generated data.
\newblock {\em Nature}, 631(8022):755--759, 2024.

\bibitem{modelsim}
{Siemens Software}.
\newblock Modelsim.

\bibitem{pyverilog}
S.~Takamaeda-Yamazaki.
\newblock Pyverilog: A python-based hardware design processing toolkit for verilog hdl.
\newblock In K.~Sano, D.~Soudris, M.~H{\"u}bner, and P.~C. Diniz, editors, {\em Applied Reconfigurable Computing}, pages 451--460, Cham, 2015. Springer International Publishing.

\bibitem{thakur}
S.~Thakur, B.~Ahmad, Z.~Fan, H.~Pearce, B.~Tan, R.~Karri, B.~Dolan-Gavitt, and S.~Garg.
\newblock Benchmarking large language models for automated verilog rtl code generation.
\newblock In {\em 2023 Design, Automation Test in Europe Conference Exhibition (DATE)}, pages 1--6, 2023.

\bibitem{thakur2024verigen}
S.~Thakur, B.~Ahmad, H.~Pearce, B.~Tan, B.~Dolan-Gavitt, R.~Karri, and S.~Garg.
\newblock Verigen: A large language model for verilog code generation.
\newblock {\em ACM Transactions on Design Automation of Electronic Systems}, 29(3):1--31, 2024.

\bibitem{instructfinetune}
J.~Wei, M.~Bosma, V.~Y. Zhao, K.~Guu, A.~W. Yu, B.~Lester, N.~Du, A.~M. Dai, and Q.~V. Le.
\newblock Finetuned language models are zero-shot learners.
\newblock {\em arXiv preprint arXiv:2109.01652}, 2021.

\bibitem{zhang2024mg}
Y.~Zhang, Z.~Yu, Y.~Fu, C.~Wan, and Y.~C. Lin.
\newblock Mg-verilog: Multi-grained dataset towards enhanced llm-assisted verilog generation.
\newblock In {\em 2024 IEEE LLM Aided Design Workshop (LAD)}, pages 1--5. IEEE, 2024.

\bibitem{zhao2024codev}
Y.~Zhao, D.~Huang, C.~Li, P.~Jin, Z.~Nan, T.~Ma, L.~Qi, Y.~Pan, Z.~Zhang, R.~Zhang, et~al.
\newblock Codev: Empowering llms for verilog generation through multi-level summarization.
\newblock {\em arXiv preprint arXiv:2407.10424}, 2024.

\bibitem{hfrl1}
D.~M. Ziegler, N.~Stiennon, J.~Wu, T.~B. Brown, A.~Radford, D.~Amodei, P.~Christiano, and G.~Irving.
\newblock Fine-tuning language models from human preferences.
\newblock {\em arXiv preprint arXiv:1909.08593}, 2019.

\end{thebibliography}

\end{document}